\documentclass[cits]{PoS}
\usepackage{multicol}

\title{A highly collimated jet from the low mass proto-star NGC1333 IRAS\,4B}

\ShortTitle{A highly collimated jet from IRAS\,4B}

\author{\speaker{J.-F. Desmurs}\\%\thanks{}\\
        Observatorio Astron\'omico Nacional, Spain\\
        E-mail: \email{desmurs@oan.es}}
\author{C. Codella\\%\thanks{}\thanks{}\\
        IRA, INAF, Italy \\
        E-mail: \email{codella@arcetri.astro.it}}
\author{J. Santiago-Garc\'{\i}a\\%\thanks{}\thanks{}\\
        Observatorio Astron\'omico Nacional, Spain\\
        E-mail: \email{j.santiago@oan.es}}
\author{M. Tafalla\\%\thanks{}\thanks{}\\
        Observatorio Astron\'omico Nacional, Spain\\
        E-mail: \email{tafalla@oan.es}}
\author{R. Bachiller\\% \thanks{}\\
        Observatorio Astron\'omico Nacional, Spain\\
        E-mail: \email{r.bachiller@oan.es}}

\abstract{We imaged the protostars of the nearby region NGC1333 IRAS 4
in the water maser line at 22.2 GHz by using the VLBA in phase
referencing at milliarcsecond scale over four epochs spaced by one
month to measure proper motions. We measure the absolute positions and
proper motions of the H$_2$O spots to investigate the kinematics of the
region from where the jet is launched.

Two protostars (A2 and B) have been detected in a highly variable
H$_2$O maser emission, with an active phase shorter than four weeks. A
70 AU chain of several maser spots, very well aligned, has been
observed close to the B protostar. The apparent proper motions have
been derived, finding that the H$_2$O spots are moving along the N-NW
direction with projected velocities between 10 and 50 km\,s$^{-1}$.  We
conclude that in IRAS\,4B, water maser trace a highly collimated
bipolar jet clearly associated with the protostar.}

\FullConference{The 9th European VLBI Network Symposium on The role of
                VLBI in the Golden Age for Radio Astronomy and EVN
                Users Meeting\\ 
                September 23-26, 2008\\ 
                Bologna, Italy}

\begin{document}

\section{Introduction} 

NGC1333 is a well-known star forming region containing a large number
of proto-stars and located at a distance of 235 pc, according to the
recent observations \cite{Hirota08} performed with VERA (VLBI
Exploration of Radio Astrometry).  The IRAS\,4 region contains three
star forming sites, called 4A, 4B, and 4C, which can be identified in
the mm-continuum.  High angular resolution cm- and mm-observations
\cite{Reipurth02} have revealed that IRAS\,4B is also associated with a
bipolar outflow located along the North-South direction
\cite{jorgensen07}.  Water (H$_2$O) masers at 22 GHz in combination
with the high angular resolution given by Very Long Baseline
Interferometry (VLBI) techniques, represent a unique tool to
investigate the mechanism of jet formation and collimation in the
earliest star forming phases.  All these characteristics make of IRAS4
an excellent laboratory where to study multiple star formation.

\section{Observations and Results} 

Using the VLBA network, we observed H$_2$O maser emission at 22.2 GHz
in phase referencing over four epochs spaced by about four weeks
during 2003 (Apr 1st, May 14th, June 11th, July 07th).  The duration
of each observations was 10 hours in total.

As the four proto-stars (IRAS\,4A1, 4A2, 4B, and 4C) fell inside the
primary beam of the antennas, during the observation we tracked their
barycenter.  Data were recorded in dual circular polarization with a
velocity resolution (i.e. channel width) of $\sim$0.1~km\,s$^{-1}$ and
a total velocity coverage of about 55~km\,s$^{-1}$.  To be able to
measure and compare the relative position of the water maser emission
across the epochs, we used the phase referencing technique with the
close phase calibrator 0333+321 (1.5 degree away). Data were correlated
in three separate passes, one per source, at the VLBA correlator in
Socorro (New Mexico, USA) with an integration time of 2 seconds. The
data reduction was performed using the Astronomical Image Processing
System (AIPS) package. The amplitude was calibrated using the template
spectra method. The final restoring beam was 1.2 $\times$ 1.2 mas and
each channel was cleaned down to a residual rms noise of $\sim$ 15 mJy.

In the first epoch, no emission has been detected towards the four
sources.  The proto-stars IRAS 4A1+4A2 and IRAS 4B have shown water
emission in two of the four epochs observed. This confirms the high
variability of the H$_2$O spots associated with low-mass young stellar
objects \cite{Wilking94, Claussen96}, with an active phase shorter than
a few weeks.

Figure 1 reports the July 07th map for IRAS\,4B with two groups of
H$_2$O spots mapped: redshited emission at NW $\sim$[+12/+17.0]
km\,s$^{-1}$, blueshited at SE $\sim$0km\,s$^{-1}$ (ambient emission is
at $V_{\rm LSR}$ = +7.25 km\,s$^{-1}$).  The size of the water spots
chain is about 270 mas ($\sim$60 AU at 235 pc) and its position angle
is P.A. $\sim$ $25^{\circ}$ NW.  As a reference, the dashed circle in
Fig. 1 reports the coordinates of IRAS\,B following \cite{Reipurth02}
who gave the position of the NGC1333 proto-stars, observed at 3.6 cm,
with a 50 mas uncertainty (size of the circle). The closest spot to
IRAS\,B, being detected at the level of the uncertainty area, is at
$\sim$ 10 AU.  The very good alignment of the maser spots as well as
their velocities suggest their association with a molecular jet driven
by the IRAS\,B proto-star and having a high degree of collimation (this
is consistent with EVN observations of this source by
\cite{Desmurs00}).  If we conservatively assume an inclination angle
$\simeq$ 20$^{\circ}$--30$^{\circ}$, we derive that the water masers
are moving towards the N-NW (Red) and the S-SE (Blue) direction with
velocities $V_{\rm H_2O}$ $\simeq$ 20--150 km s$^{-1}$, in agreement
with the high velocities usually observed along jets form proto-stars.

Our observations are consistent with those recently published by
\cite{Marvel08} using VLBA observations from 1998 from water masers
associated with NGC 1333--IRAS4B. The orientation of the jet traced by
the water maser emission has nearly the same position angle
(observations of the two projects were separated by five years), and
shows high velocities but does not seems to coincide with the large
scale outflow located along the N--S direction.

\begin {figure}
\begin{minipage}[]{0.5\textwidth}
   \includegraphics[angle=0,width=\textwidth]{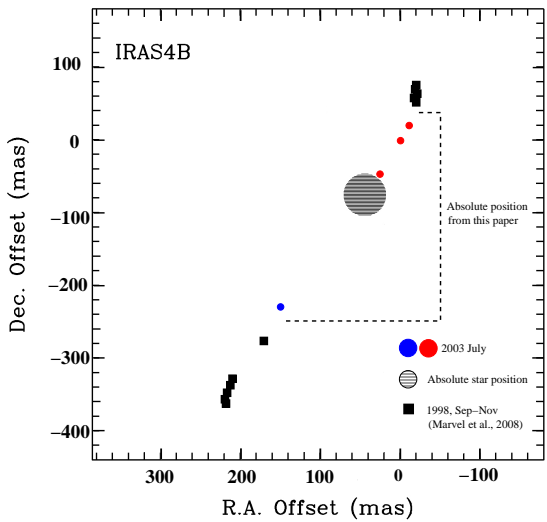}
\end{minipage}
\hspace{0.7cm}
\begin{minipage}[]{0.4\textwidth}
 \caption{Figure 1. Water maser maps measured for IRAS\,4B (July 07th).
The map reports the chain of H$_2$O spots distributed over $\sim$70
AU.  The driving proto-star (hash circle; \cite{Reipurth02}) is at
+66,--95 mas with respect to the center of the present maps, at
$\alpha_{\rm 2000}$ = 03$^{\rm h}$ 29$^{\rm m}$ 12$^{\rm s}$003,
$\delta_{\rm 2000}$ = +31$^{o}$ 13$'$ 08$''$14.  The size of the circle
stands for the uncertainty (50 mas) associated with the position of the
proto-star. Black square, report the water masers observed by
\cite{Marvel08}, their locations are just indicative with respect to
our results.}
 \label{fig}
\end{minipage}
\end{figure}

Further observations at high angular resolution of the molecular
outflow are needed to clarify whether the disagreement between the
molecular outflow and the water jet is due to precession or whether we
simply observe different components driven by different sources.

\end{document}